\documentstyle[aps,prl,epsf]{revtex}
\begin{document}
\draft
\title{Collective modes and sound propagation in a $p$-wave 
superconductor: Sr$_2$RuO$_4$}

\author{Hae-Young Kee$^{1}$, Yong Baek Kim$^{2}$, and Kazumi Maki$^{3}$}
\address{
$^1$ Department of Physics, University of California, 
Los Angeles, CA 90095\\
$^2$ Department of Physics, The Ohio State University, 
Columbus, OH 43210\\
$^3$ Department of Physics, 
University of Southern California, Los Angeles, CA 90089}

\date{\today}
\maketitle

\begin{abstract}

There are five distinct collective modes in the recently
discovered $p$-wave superconductor Sr$_2$RuO$_4$;
phase and amplitude modes of the order parameter, 
clapping mode (real and imaginary), and spin wave.
The first two modes also exist in the ordinary s-wave superconductors,
while the clapping mode with the energy $\sqrt{2} \Delta(T)$
is unique to Sr$_2$RuO$_4$ and couples to the sound wave.
Here we report a theoretical study of the sound propagation in a
two dimensional p-wave superconductor.
We identified the clapping mode and study its effects on
the longitudinal and transverse sound velocities 
in the superconducting state. 
In contrast to the case of $^3$He,
there is no resonance absorption associated with 
the collective mode, since in metals
$\omega/(v_F |{\bf q}|) \ll 1$, where $v_F$ is the Fermi
velocity, {\bf q} is the wave vector, and $\omega$ is the 
frequency of the sound wave.
However, the velocity change in the collisionless limit gets 
modified by the contribution from the coupling to 
the clapping mode. We compute this contribution and comment
on the visibility of the effect.
In the diffusive limit, the contribution from the collective
mode turns out to be negligible.
The behaviors of the sound velocity change and the attenuation
coefficient near $T_c$ in the diffusive limit are calculated
and compared with the existing experimental data wherever it
is possible. We also present the results for the attenuation 
coefficients in both of the collisionless and diffusive 
limits at finite temperatures.
\end{abstract}


\section{Introduction}

Shortly after the discovery of superconductivity in Sr$_2$RuO$_4$,
the possibility of spin triplet pairing was discussed.\cite{rice}
Possible pairing symmetries were also classified based on the
crystal symmetry.\cite{sigrist} On the experimental front, there have
been attempts to single out the right pairing symmetry among
these possibilities.
Recent measurement of $^{17}$O-Knight shift in NMR for the 
magnetic field parallel to the $a-b$ plane showed no change 
across $T_c$, which can be taken
as the evidence of the spin triplet pairing with 
${\hat d}$-vector 
parallel to the $c$-axis.\cite{ishida}
Here ${\hat d}$ is called the spin vector which
is perpendicular to the direction of the spin associated with the 
condensed pair\cite{vollhardt}.
$\mu$SR experiment found spontaneous magnetic field in the 
superconducting Sr$_2$RuO$_4$, which seems to indicate broken time reversal 
symmetry in the superconducting state. \cite{luke}
These experiment may be compatible with the following 
order parameter \cite{sigrist}
\begin{equation}
{\hat \Delta}({\bf k})=\Delta {\hat d} (k_1 \pm i k_2),
\end{equation}
where
$\Delta$ is the magnitude of the superconducting order parameter.
Notice that this state is analogous to the $A$ phase of $^3$He
and there is a full gap on the Fermi surface.

On the other hand, there also exist experiments that cannot be
explained by a naive application of the order parameter given
by Eq.(1). Earlier specific heat measurement found residual 
density of states at low temperatures below $T_c$\cite{nishizaki}, 
which provokes
the ideas of orbital dependent superconductivity\cite{agterberg}
and even non-unitary superconducting state\cite{sigrist3}.
However, more recent specific heat experiment on a cleaner sample
reports no residual density of states and it was found that
the specific heat behaves as $T^2$ at low temperatures.
\cite{nishizaki2}
This result stimulated a speculation about different order parameters 
with line node\cite{hasegawa}.
However, since there are three bands labeled by $\alpha$, $\beta$, 
and $\gamma$ which cross the Fermi surface,
it is not yet clear whether the order parameter given by Eq.(1)
is compatible with more recent specific heat data or not.
For example, it is possible that the pairing symmetry associated
with the $\gamma$ band is still given by Eq.(1) while the order
parameter symmetry associated with $\alpha$ and $\beta$ bands
can be quite different. In this case, the low temperature 
specific heat will be dominated by the excitations from 
$\alpha$ and $\beta$ bands. In order to resolve the issue, it is
important to examine other predictions of the given order parameter 
and compare the results with future experiments. 

One way of identifying the correct order parameter among possible
candidates is to investigate the unique collective modes supported by
the ground state with a given pairing symmetry.
The observation of the effects of these collective modes
would provide convincing evidence for a particular 
order parameter symmetry.
If we assume that the order parameter of Eq.(1) 
is realized in Sr$_2$RuO$_4$, the superconducting state
would support unique collective modes, the so-called clapping mode 
and spin waves as well as the phase and amplitude modes of
the order parameter which exist also in $s$-wave
superconductors.    
Previously we studied the dynamics of spin waves \cite{hykee,tedwort}.
A possible way to distinguish the order parameter of the
$\gamma$ band from those of the $\alpha$ (or $\beta$) band 
was also proposed in the context of spin wave dynamics.\cite{hykee2}

In this paper, we study the dynamics of the sound wave and its 
coupling to the clapping modes assuming that the order parameter
is given by Eq.(1). As in $^3$He, only the clapping mode can 
couple to the sound wave and affects its dynamics.
Here we study the sound velocities and attenuation
coefficients of the longitudinal and transverse sound waves. 
In particular, we identify the clapping mode 
with the frequency, $\omega = \sqrt{2} \Delta (T)$, and examine
the effects of this mode and disorder on the sound wave
propagation.

In a recent paper, Higashitani and Nagai\cite{nagai} obtained
the clapping mode with the frequency, $\omega=\sqrt{2} \Delta$, and 
discussed the possible coupling to the sound wave independent of us.
It is, however, important to realize that the coupling 
to the sound wave is extermely small because $C/v_F \ll 1$ in metals, 
where $C$ is the sound velocity.
Indeed the recent measurement of the sound velocity in the normal 
and superconducting states of Sr$_2$RuO$_4$ reported in \cite{matsui}
shows that $C/v_F \ll 1$.
They measured the sound velocities of the longitudinal modes,
$C_{11}$ (${\bf q},{\bf u} \parallel [100]$) 
and $C_{33}$ (${\bf q},{\bf u} \parallel [001]$), 
and the transverse modes,
$C_{44}$ (${\bf q} \parallel [100]$, ${\bf u} \parallel [001]$) 
and 
$C_{66}$ (${\bf q} \parallel [100]$, ${\bf u} \parallel [010]$),
where ${\bf q}$ and ${\bf u}$ are the directions of 
propagation and the polarization of ultrasound, respectively. 
They found that the longitudinal
sound velocities, $C_{11}$ and $C_{33}$, decrease with a kink at $T=T_c$, 
while the transverse sound velocities do not exhibit any effect
of the onset of superconductivity.
We estimate from their experimental data that 
$C_l/v_F \sim 10^{-2}$, where 
$C_l$ is the longitudinal sound velocity. 
It can be also seen that the transverse sound velocity, $C_t$, 
is much smaller than the longitudinal one \cite{matsui}.

Incorporating the correct limit, $C/v_F \ll 1$, we obtained
the sound velocities and attenuation coefficients for 
both collisionless and diffusive limits. In the diffusive
limit, the quasi-particle scattering due to impurities
should be properly taken into account.
One can show that, in a metal like Sr$_2$RuO$_4$, the collisionless 
limit is rather difficult to reach because it can be 
realized only for $\omega \sim O(1)$ GHz. For more practical range
of frequencies, kHz $-$ MHz, the diffusive limit may be
easier to achieve.     
On the other hand, we found that it is much easier to see 
the effects of the coupling between the sound waves and the 
clapping mode in the collisionless limit.
Therefore it is worthwhile to study both regimes.  

Here we summarize our main results.

\vskip 0.1cm
\noindent
{\bf A. Collisionless limit}

In the absence of the coupling to the clapping mode,
the longitudinal sound velocity decreases in the
superconducting state because the effect of the screening 
of the Coulomb potential increases, which happens in the 
s-wave superconductors as well.
However, one of the important features of the $p$-wave
order parameter in consideration is that the sound wave
can now couple to the clapping mode. This effect is absent
in $s$-wave superconductors.
One can show that, among longitudinal waves, $C_{11}$ mode 
can couple to the clapping mode, but $C_{33}$ mode cannot.
We found that the longitudinal sound velocity $C_{11}$ 
decreases as
\begin{equation}
\frac{\delta C^{11}_l}{C^{11}_l} = -\lambda^{11}_l \left [ \frac{1}{2}-2
\left ({C^{11}_l \over v_F} \right )^2 (1-f-
\frac{f} {4 \{1+(2\Delta(T)/v_F q)^2\} } ) \right ] \ .
\end{equation}
where $\lambda_l$ is the couping constant and 
$f$ is the superfluid density.
$\delta C_l / C_l$ is the relative shift in the sound velocity.
We estimated the frequency regime where one can observe the 
effect of the clapping mode and found that the effect 
is visible if  $v_F|{\bf q}| \sim 2 - 3 \Delta(0)$.
This implies that $\omega \sim  O(1)$ GHz.
Since $C_{33}$ does not couple to the clapping mode, the
velocity change is simply given by
\begin{equation}
\frac{\delta C^{33}_l}{C^{33}_l} = -\lambda^{33}_l \left [ \frac{1}{2}-2
\left ({C^{33}_l \over v_F} \right )^2 (1-f) \right ] \ .
\end{equation}

Since the velocity of the transverse wave is much smaller than that
of the longitudinal one and the coupling to the electron system is
weaker than the longitudinal case as well, we expect that
the change of the transverse sound velocity is hard to observe.
In order to complete the discussion, we also present the
results for these small changes in transverse velocities.
Here only $C_{66}$ mode couples to the clapping mode, and
$C_{44}$ mode does not. We found
\begin{equation}
\frac{\delta C^{66}_t}{C^{66}_t} = -\lambda^{66}_t 
\left [ \frac{1}{2}+
2 \left ( {C^{66}_t \over v_F} \right )^2 \left( 1-f+
\frac{f}{4 \{ 1+(2\Delta/v_F q)^2\} } \right ) \right ] \ ,
\end{equation}
\begin{equation}
\frac{\delta C^{44}_t}{C^{44}_t} = -\lambda^{44}_t 
\left[ \frac{1}{2}+
2 \left ( \frac{C^{44}_t}{v_F} \right )^2 (1-f)  \right] \ ,
\end{equation}
where the $\lambda_t$ is the transverse coupling 
constant.

We also found that the leading contribution to the attenuation 
coefficient is the same as that of $s$-wave superconductors
in the collisionless limit. 

\vskip 0.1cm
\noindent
{\bf B. Diffusive limit}

This case corresponds to $\omega, v_F|{\bf q}| \ll \Gamma$, where
$\Gamma$ is the scattering rate due to impurities.
As in the case of the collisionless limit, in principle 
$C_{11}$ and $C_{66}$ modes
couple to the clapping mode, but it turns out that the effect is
almost impossible to detect.
Neglecting the coupling to the clapping mode and working in 
the limit $4 \pi T_c \gg 2\Gamma \gg v_F |{\bf q}|, \omega$,
we obtain the following results near $T_c$. 
\begin{eqnarray}
\frac{\delta C_l}{C_l} &=& -\lambda_{l} 
{1 \over 2} \left \{
1 - \left ( \frac{\omega}{2\Gamma} \right )^2
\left [ 1-\frac{4 \pi^3}{7 \zeta(3)} \frac{T}{\Gamma}
\left ( 1-\frac{T}{T_c} \right ) \right ] \right \} \ ,
\nonumber\\
\frac{\alpha_l}{\alpha^n_l} &=&
1-\frac{2 \pi^3}{7 \zeta(3)} \frac{T}{\Gamma} 
\left ( 1-\frac{T}{T_c} \right) \ ,
\label{diffusive}
\end{eqnarray}
where $\alpha$ and $\alpha_n$ are the attenuation coefficients
in the superconducting and the normal states respectively.
The shift of the sound velocity and attenuation coefficient
decrease linearly in $(1-T/T_c)$ as $T \rightarrow T_c$.
This result for the longitudinal sound wave is consistent with 
the experimental observation reported in \cite{matsui}

In the case of the transverse sound waves, the leading 
behaviors of the sound velocity and the attenuation coeffient 
can be obtained simply by replacing $C_l$ and $\lambda_l$ by
$C_t$ and $\lambda_t$ in the diffusive limit.  
However, the absolute value of the transverse sound velocity is 
much smaller than the longitudinal one and 
the coupling to the electron system is also much weaker than
the case of the longitudinal sound waves.
Thus, it would be hard to observe any change at $T=T_c$ for
the transverse wave.
This may explain the experimental finding that
the transverse velocity does not show any change 
across $T_c$.
We also obtained the attenuation coefficient for all 
temperatures below $T_c$. It is given by
Eq.(\ref{att}) and Fig.2 shows its behavior.  
 
The rest of the paper is organized as follows.
In section II, the clapping mode is briefly discussed.
In section III, we provide a brief summary of the formalism 
used in Ref.\cite{kadanoff} to explain how the sound 
velocity and the attenuation coefficients are related to 
the autocorrelation functions of the stress tensor.
We present the results of the study on the sound 
propagation in the collisionless and diffusive
limits in sections IV and V, respectively.
We conclude in section VI.
Further details which are not presented in the main
text are relegated to the Appendix A and B.  

\section{Collective modes in Sr$_2$RuO$_4$}

As in the $s$-wave superconductors, the phase and amplitude
modes of the order parameter also exist in the $p$-wave
superconductors. On the other hand, due to the internal structure
of the Cooper pair in the $p$-wave superconductor, there
exist other types of collective mode associated with the
order parameter. The nature of these modes is determined
by the structure of the order parameter.

There are collective modes associated with the oscillation of the
spin vector ${\hat d}$, which we have already discussed
in \cite{hykee} and \cite{tedwort}.
There exists another collective mode associated with the orbital part.
Using the notation $e^{\pm i \phi}=(k_1 \pm i k_2)/|{\bf k}|$, 
the oscillation of the orbital part $e^{\pm i \phi} \rightarrow
e^{\mp i \phi}$ gives rise to the clapping mode with 
$\omega=\sqrt{2}\Delta(T)$. This mode couples to the sound waves
as we will see in the next section.
Therefore, the detection of the clapping mode will provide a
unique evidence for the $p$-wave superconducting order parameter.
The derivation of the clapping mode and the coupling to 
the sound wave is discussed in Appendix A.

\section{Dynamics of sound wave via stress tensor}

In ordinary liquids, the sound wave is a density wave.
In superconductors, the density is not only coupled
to the longitudinal component of the normal velocity,
but also to the superfluid velocity and to temperature.
The role of these couplings and their consequences
in the dynamics of sound wave can be studied by looking
at the autocorrelation function,
$\langle [\tau_{ij}, \tau_{ij}] \rangle$,
of stress tensor,
$\tau_{ij}$.
\begin{equation}
\langle [\tau_{ij}, \tau_{ij}] \rangle ({\bf r}-{\bf r'}, t-t')
\equiv - i \theta (t-t')
\langle [\tau_{ij} ({\bf r},t), \tau_{ij} ({\bf r'},t') ]
\rangle \ ,
\end{equation}
where
\begin{equation}
\tau_{ij} ({\bf r},t) = \sum_{\sigma}
\left [
{(\nabla-\nabla^{'})_i \over 2i}{(\nabla-\nabla^{'})_j \over
2im} \psi^{\dagger}_{\sigma} ({\bf r},t)
\psi_{\sigma} ({\bf r'},t)
\right ]_{{\bf r'}={\bf r}} \ .
\end{equation}
Here $\psi^{\dagger}_{\sigma}$ is the electron creation
operator with spin $\sigma$.

The other operators whose correlation functions are needed
for the ultrasonic attenuation and the sound velocity change
are the density operator
\begin{equation}
n({\bf r},t)= \sum_{\sigma} \psi^{\dagger}_{\sigma}({\bf r},t)
\psi_{\sigma}({\bf r},t),
\end{equation}
and the current operator
\begin{equation}
{\bf j}({\bf r},t)=\sum_{\sigma}
\left [
{(\nabla-\nabla^{'})_j \over 2im} \psi^{\dagger}_{\sigma} ({\bf r},t)
\psi_{\sigma} ({\bf r'},t) \right ]_{{\bf r'}={\bf r}} \ .
\end{equation}

Assuming that the wave vector of the sound wave
is in the ${\hat {\bf x}}$ direction, ${\bf q}=q {\hat x}$,
the sound velocity shift,
$\delta C$, at low frequencies can be computed from
\begin{eqnarray}
{\delta C_l \over C_l} &=& \left. \frac{C_l(\omega)-C_l}{C_l} \right
|_{\omega=C_l |{\bf q}|}
= - \left. {\omega  \over  m_{\rm ion} C_l |{\bf q}|}
{\rm Re} \langle [h_l,h_l] \rangle
%{\rm Re} \langle [\tau_{xx}, \tau_{xx}] \rangle
({\bf q},\omega) \right |_{\omega=C_l |{\bf q}|} \ ,
\nonumber\\
{\delta C_t \over C_t} &=& \left. \frac{C_t(\omega)-C_t}{C_t} \right
|_{\omega=C_t |{\bf q}|}
= - \left. {\omega  \over  m_{\rm ion} C_t |{\bf q}|}
{\rm Re} \langle [h_t,h_t] \rangle
%{\rm Re} \langle [\tau_{xy}, \tau_{xy}] \rangle
({\bf q},\omega) \right |_{\omega=C_t|{\bf q}|} \ ,
\end{eqnarray}
where
\begin{eqnarray}
h_l({\bf r},t)& =& \frac{q}{\omega} \tau_{xx}({\bf q},t)-
\frac{\omega m}{q} n({\bf r},t),
\nonumber\\
h_t({\bf r},t) & =& \frac{q}{\omega} \tau_{xy}({\bf q},t)-
 m j_y({\bf r},t),
\end{eqnarray}
Here $C_l$ and $C_t$ represent the longitudinal and transverse
sound velocities in the normal state respectively.
$m_{\rm ion}$ and $m$ are the mass of ions and the mass of electron,
respectively.
On the other hand, the attenuation coefficient, $\alpha$,
at low frequencies is obtained from
\begin{equation}
\alpha_l = \left. {\omega \over m_{\rm ion} C_l}
{\rm Im} \langle [h_l,h_l] \rangle
%{|{\bf q}|^2 \over \omega}
%{\rm Im} \langle [\tau_{xx}, \tau_{xx}] \rangle
({\bf q},\omega) \right |_{\omega=C_l |{\bf q}|} \ ,
\hskip 0.5cm
\alpha_t =  \left. {\omega \over m_{\rm ion} C_t}
%{|{\bf q}|^2 \over \omega}
%{\rm Im} \langle [\tau_{xy}, \tau_{xy}] \rangle
{\rm Im} \langle [h_t,h_t] \rangle
({\bf q},\omega) \right |_{\omega=C_t |{\bf q}|} \ .
\end{equation}
These relations are extensively discussed in the work of
Kadanoff and Falko.\cite{kadanoff}

\section{Sound propagation in the collisionless limit}

As discussed in the introduction, in a metal like Sr$_2$RuO$_4$ 
the collisionless limit is somewhat difficult to reach because 
we need the sound wave with the frequency $\omega \sim O(1)$ GHz.
However, we will also see that this is the regime where one
has the best chance to observe the effect of the collective mode.

The sound velocity shift and the attenuation coefficients 
can be calculated by looking at the autocorrelation functions, 
$\langle [\tau_{ij}, \tau_{ij}] \rangle$,
of stress tensor, $\tau_{ij}$, as discussed in the previous section.
We will use the finite temperature Green's function technique\cite{maki} 
to compute these correlation functions. 
The single particle Green's function, $G (i\omega_n,{\bf k})$, 
in the Nambu space is given by
\begin{equation}
G^{-1}(i\omega_n,{\bf k})=i \omega_n - \xi_{\bf k}\rho_3 
- \Delta ({\hat k}\cdot {\hat \rho}) \sigma_1,
\label{green}
\end{equation}
where $\rho_i$ and $\sigma_i$ are Pauli matrices acting on
the particle-hole and spin space respectively,
$\omega_n = (2n+1) \pi T$ is the fermionic Matsubara 
frequency, and $\xi_{\bf k}={\bf k}^2/2m - \mu$.
Then, for example, the irreducible correlation function can be 
computed from
\begin{equation}
\langle [\tau_{ij},\tau_{ij}] \rangle_{00} (i\omega_{\nu},{\bf q})
=T \sum_n \sum_{\bf p} {\rm Tr} [\left ({p_i p_j \over m} \right )^2
\rho_3 G({\bf p},\omega_n)
\rho_3 G({\bf p}-{\bf q},i\omega_n-i\omega_{\nu})] \ ,
\label{irr}
\end{equation}
where 
$\omega_{\nu} = 2\nu \pi T$ is the bosonic Matsubara frequency.

\subsection{Longitudinal sound wave}

Let us consider the longitudinal wave with ${\bf u} \parallel
{\bf q} \parallel {\bf x}$, which corresponds to the $C_{11}$ mode.
Since the stress tensor couples to the density, the autocorrelation 
function $\langle [h_l, h_l] \rangle$ is renormalized by the 
Coulomb interaction. In the long wavelength limit and for 
$s \equiv \omega / v_F |{\bf q}| \ll 1$, the renormalized
correlation function $\langle [h_l, h_l] \rangle_0$ can be 
reduced to
\begin{equation}
{\rm Re} \langle [h_l,h_l] \rangle_{0} \approx 
\left( \frac{q}{\omega} \right)^2 {\rm Re} {p^4_F \over 4 m^2}
\langle {\rm cos}(2\phi), {\rm cos}(2\phi) \rangle
= {p^4_F \over 4 m^2} N(0) \left [ 
\frac{1}{2}-2 s^2 (1-f) 
\right ] \ ,
\label{hlhl}
\end{equation}
where
\begin{equation}
\langle A, B \rangle = 
T \sum_n \sum_{\bf p} {\rm Tr} [ A 
\rho_3 G({\bf p},\omega_n) 
B \rho_3 G({\bf p}-{\bf q},i\omega_n-i\omega_{\nu})] \ ,
\end{equation}
with $A$ and $B$ being some functions of $\phi$ or
operators. Here $\phi$ is the angle between {\bf p}
and {\bf q}, $N(0)=m/2\pi$ is the density of states 
at the Fermi level and 
$f$ is the superfluid density in the static limit
($\omega \ll v_F|{\bf q}|$) given by
\begin{equation}
f= 2\pi T \Delta^2 \sum^{\infty}_{n=0} \frac{1}{\omega_n^2+\Delta^2}
\frac{1}{\sqrt{\omega_n^2+\Delta^2+(v_F q)^2/4}} \ .
\label{f}
\end{equation}
The derivation of the result in Eq.(\ref{hlhl}) is given
in Appendix B 2. 
Therefore, the sound velocity shift, $\delta C_{l}$, is given by 
\begin{equation}
\frac{\delta C_l}{C_l} = -\lambda_l [\frac{1}{2}-2 s^2 (1-f) ],
\end{equation}
where $\lambda_l = p^2_F/(8 \pi m m_{\rm ion} C^2_l)$ is the 
longitudinal coupling constant.
Here we set $s=\omega/v_F|{\bf q}|= C_l/v_F$, where $C_l$ is the 
longitudinal sound velocity. 
In Sr$_2$RuO$_4$, $s=10^{-2} \ll 1$ which is very different 
from $s \gg 1$ of $^3$He.

Now let us consider the correction due to the collective modes.
The additional renormalization of 
$\langle [h_l, h_l] \rangle$ (in the $C_{11}$ mode)
due to the collective mode is 
computed in Appendix B 3 and the result is given by
\begin{equation}
\langle [h_l,h_l] \rangle =
\left ( {q \over \omega} \right )^2 {p^4_F \over 4 m^2} \left [
\frac{1}{2}-2 s^2 \left ( 1-f-
\frac{f (v_F |{\bf q}|)^2}
{4 \{(v_F|{\bf q}|)^2+4\Delta(T)^2-2\omega^2\} } \right ) \right ]
+i {m_{\rm ion} C^{11}_l \over \omega} \alpha_l (\omega) \ .
\end{equation}
As one can see from the above equation, there is no resonance
because $\omega \ll v_F |{\bf q}|$.
However, we will be able to see a shadow of the 
collective mode in the sound velocity change, which we discuss
in the following. 

In the limit $s \ll 1$ and setting $s=C^{11}_l/v_F$, 
the above equation leads to the sound velocity shift 
given by
\begin{equation}
\frac{\delta C^{11}_l}{C^{11}_l} = -\lambda^{11}_l \left [ 
\frac{1}{2}-2 
\left ({C^{11}_l \over v_F} \right )^2 \left ( 1-f-
\frac{f} {4 \{1+(2\Delta/v_F q)^2\} } \right ) \right ] \ .
\end{equation}
Note that the sound wave gets soften more by the collective mode.
In Fig. 1, we show $I=1-f-\frac{f}{4[1+(2\Delta(T)/v_F|{\bf q}|)^2]}$
for $v_F|{\bf q}|/ \Delta (0)=0,1,2,3$ for $0.7 < t < 1.0$ where
$t=T/T_c$.
Note that the coupling to the collective mode can be observed
for $v_F|{\bf q}| \sim 2-3\Delta(0)$
which corresponds to $\omega \sim O(1) $ GHz.

The attenuation coefficient, $\alpha_l$, is given by 
\begin{eqnarray}
{\alpha_l (\omega) \over \alpha^n_l (\omega)} &=&\frac{1}{\omega} 
\int^{\infty}_{\Delta} 
d\omega^{\prime} \frac{\omega^{\prime}
(\omega^{\prime}+\omega)-\Delta^2}
{\sqrt{\omega^{\prime 2}-\Delta^2} \sqrt{(\omega^{\prime}+\omega)^2-\Delta^2}}
\left(
\tanh\frac{\omega+\omega^{\prime}}{2T}-\tanh\frac{\omega^{\prime}}{2T} \right)
\nonumber\\
&- & 
\theta(\omega-2\Delta) \frac{1}{\omega} \int^{\omega-\Delta}_{\Delta}
d\omega^{\prime} \frac{\omega^{\prime}
(\omega^{\prime}-\omega)-\Delta^2}
{\sqrt{\omega^{\prime 2}-\Delta^2} \sqrt{(\omega^{\prime}-\omega)^2-\Delta^2}}
\left(
\tanh\frac{\omega^{\prime}}{2T} \right),
\end{eqnarray}
where $\alpha^n_l$ is the attenuation coefficient in the normal
state. This form is the same as the one in the s-wave 
superconductors.

We can carry out a parallel analysis for the longitudinal wave with
${\bf u} \parallel {\bf q} \parallel {\bf z}$ which corresponds to 
the $C_{33}$ mode.
Unfortunately, this sound wave does not couple to the 
clapping mode.
Therefore, the velocity shift of this sound wave is simply 
given by
\begin{equation}
\frac{\delta C^{33}_l}{C^{33}_l}=-\lambda^{33}_l
\left [ \frac{1}{2}-2 
\left ( \frac{C_l^{33}}{v_F} \right )^2 (1-f) \right] \ .
\end{equation}.

\subsection{Transverse sound wave}

Here we consider first the $C_{66}$ mode that has
${\bf u} \parallel {\bf y}$ and ${\bf q}\parallel {\bf x}$.
In this case, the sound velocity change can be obtained from 
the evaluation of $\langle [h_t, h_t] \rangle$. 
Assuming that the current contribution is negligible at
low frequencies and  
following the same procedure used in the case of the
longitudinal sound wave, we obtain
\begin{equation}
\frac{\delta C^{66}_t}{C^{66}_t} = -\lambda^{66}_t 
\left [ \frac{1}{2}+
2 \left ( {C^{66}_t \over v_F} \right )^2 \left( 1-f+
\frac{f}{4 \{ 1+(2\Delta/v_F q)^2\} } \right ) \right ],
\end{equation}
where the $\lambda_t = 
p^2_F/(8 \pi m m_{\rm ion} C^2_t)$ 
is the transverse coupling constant.
Note that the transverse sound velocity increases 
upon entering the superconducting state.
However, due to the fact that the transverse velocity is rather small
and the coupling to the electron system is also weak compared to the
longitudinal case, it will be hard to observe the change of 
the transverse sound velocity at $T=T_c$.

Another transverse sound mode, $C_{44}$, that 
has ${\bf u} \parallel {\bf z}$ and ${\bf q} \parallel {\bf x}$, 
does not couple to the clapping mode. Thus the sound velocity 
change in this case is given by
\begin{equation}
\frac{\delta C^{44}_t}{C^{44}_t} = -\lambda^{44}_t 
\left[ \frac{1}{2}+
2 \left ( \frac{C^{44}_t}{v_F} \right )^2(1-f)  \right] \ .
\end{equation}

\section{The diffusive limit}

In the frequency range kHz $\sim$ MHz, the diffusive limit is more
realistic.
In this limit, the incorporation of the quasi-particle damping is 
very important.
Here we assume for simplicity that the quasi-particle scattering 
is due to impurities.
Unlike the case of s-wave superconductors, we treat the impurity 
scattering in the unitary limit.
Then the effect of the impurity is incorporated by changing 
$\omega_n$ to ${\tilde \omega}_n$ (renormalized Matsubara frequnecy) 
in Eq.(\ref{green}) \cite{maki2}.
The impurity renormalized complex frequency, ${\tilde \omega}_n$
is determined from
\begin{equation}
{\tilde \omega}_n = \omega + \Gamma 
\frac{\sqrt{{\tilde \omega}_n^2+\Delta^2}}
{{\tilde \omega}_n},
\end{equation}
where $\Gamma$ is the quasi-particle scattering rate
and the quasi-particle mean free path is given by
$l = v_F/(2\Gamma)$.

In order to compare the results in the normal state
and those in the superconducting state, let us first 
work out the correlation functions in the normal state, 
where $\Delta=0$.
   
\subsection{Normal state}

We can use Eq.(\ref{sv2}) to compute 
$\langle [h_l,h_l] \rangle_0$.
In the limit of $\omega, v_F q \ll 2 \Gamma$, we get 
\begin{equation}
\langle {\rm cos}(2\phi),{\rm cos}(2\phi) \rangle
=\langle {\rm cos}^2(2\phi) 
\left( 1-\frac{\omega}{\omega+2i \Gamma-\zeta}
\right) \rangle
\approx 
\frac{1}{2}\left 
[1- (\frac{\omega}{2\Gamma})^2+i\frac{\omega}{2\Gamma} \right ] \ .
\end{equation}
One can show that 
$\langle {\rm cos}(2\phi),1 \rangle$ is of higher 
order in $\omega/2\Gamma$ and $v_F|{\bf q}|/2\Gamma$
while $\langle 1,1 \rangle \approx 2 
\langle {\rm cos}(2\phi),{\rm cos}(2\phi) \rangle$
to the lowest order. Thus, as in the previous section,
$\langle [h_l,h_l] \rangle$ is well approximated
by $\langle {\rm cos}(2\phi),{\rm cos}(2\phi) \rangle$
times a multiplicative factor.
This gives us
\begin{equation}
{\delta C^n_l \over C_l} \approx - \lambda_l 
\frac{1}{2}\left [1 - (\frac{\omega}{2\Gamma})^2 \right ] \ ,
\hskip 0.3cm 
\alpha^n_l \approx \lambda_l |{\bf q}| 
(\frac{\omega}{2\Gamma}) \ ,
\end{equation}
where $\omega$ is set to $C_l |{\bf q}|$.

It is not difficult to see that the results of the 
transverse sound wave is essentially the same
as the longitudinal case up to the lowest order
with a simple replacement of $\lambda_l$ and
$C_l$ by $\lambda_t$ and $C_t$.
Therefore, in the diffusive limit, the longitudinal and 
transverse sound velocities have the same form
with different coupling constants.

\subsection{Superconducting state near $T_c$}

Now we turn to the case of the superconducting state near $T_c$,
where the correlation functions can be computed from Eq.(\ref{tautau})
after replacing $\omega_n$ by ${\tilde \omega}_n$.
In this section, we will assume $4\pi T_c \gg 2\Gamma 
\gg v_F |{\bf q}|$ and 
use ${\Delta \over 2 \pi T} \ll 1$ near $T_c$. 
As in the previous sections, the leading contribution 
in $\langle [h_l,h_l] \rangle$
can be computed from 
$\langle {\rm cos}(2\phi),{\rm cos}(2\phi) \rangle$.  

After some algebra, we finally obtain
\begin{eqnarray}
\langle {\rm cos}(2\phi),{\rm cos}(2\phi) \rangle
&\approx& {1 \over 2}  - {1 \over 2}  
(\frac{\omega}{2\Gamma})^2 \left [ 
1-\frac{\Delta^2}{\pi \Gamma T}
\left \{ \psi^{(1)}(\frac{1}{2}+\frac{\Gamma}{2\pi T})
-\frac{\Gamma}{8\pi T} \psi^{(2)}(\frac{1}{2}+\frac{\Gamma}{2\pi T}) 
\right \}
\right ]  \cr
&&+ i{1 \over 2}(\frac{\omega}{2\Gamma})
\left [ 1-\frac{\Delta^2}{ 2\pi \Gamma T}
\left \{ \psi^{(1)}(\frac{1}{2}+\frac{\Gamma}{2\pi T})
-\frac{\Gamma}{4\pi T} \psi^{(2)}(\frac{1}{2}+\frac{\Gamma}{2\pi T}) 
\right \} \right ] \ ,
\end{eqnarray}
where
\begin{equation}
\psi^{(n)}(z) = \left ( {d \over dz} \right )^n \psi(z) = (-1)^{n+1} n! 
\sum_{k=0}^{\infty} {1 \over (z+k)^{n+1}} \ .  
\end{equation}
Here $\psi^{(n)}(z)$ is the poly-Gamma function and 
$\psi (z)$ is the di-Gamma function.
This leads to 
\begin{eqnarray}
\frac{\delta C_{l,t}}{C_{l,t}} &=& 
-\lambda_{l,t} {1 \over 2} \left [ 1 - 
\left( \frac{\omega}{2\Gamma} \right)^2
\left( 1-\frac{\Delta^2}{\pi \Gamma T} \psi^{(1)} (\frac{1}{2}+
\frac{\Gamma}{ 2\pi T}) \right) \right ] \ ,
\nonumber\\
\frac{\alpha_{l,t}}{\alpha^n_{l,t}} &=&
1-\frac{\Delta^2}{2 \pi \Gamma T} \psi^{(1)} (\frac{1}{2}+
\frac{\Gamma}{ 2\pi T}) \ .
\end{eqnarray}
where $\alpha_n$ is the ultrasonic attenuation coefficient
in the normal state.
Here we combined the subscripts $l$ and $t$, because the above 
analysis applies to the case of the transverse sound wave as well.
Only the coupling constants $\lambda_{l,t}$ are different.
In particular, when $\Gamma/ 2 \pi T \approx \Gamma/ 2 \pi T_c \ll 1$,
the above equations can be further reduced to
\begin{eqnarray}
\frac{\delta C_{l,t}}{C_{l,t}} &=& -\lambda_{l,t} 
{1 \over 2} \left \{
1 - \left ( \frac{\omega}{2\Gamma} \right )^2
\left [ 1-\frac{4 \pi^3}{7 \zeta(3)} \frac{T}{\Gamma}
\left ( 1-\frac{T}{T_c} \right ) \right ] \right \} \ ,
\nonumber\\
\frac{\alpha_{l,t}}{\alpha^n_{l,t}} &=&
1-\frac{2 \pi^3}{7 \zeta(3)} \frac{T}{\Gamma} 
\left ( 1-\frac{T}{T_c} \right) \ .
\label{final}
\end{eqnarray}

Note that the sound velocity change and the 
attenuation coefficients decrease linearly
in $(1-T/T_c)$ as $T \rightarrow T_c$.
This result, Eq. (\ref{final}),
for the longitudinal sound wave is consistent with 
the experimental observation reported in \cite{matsui}
However, in the experiment, the transverse sound velocity 
does not show any change across $T_c$.
The absolute value of the transverse sound velocity is 
much smaller than the longitudinal one and 
the coupling to the electron system is also much weaker than that 
of the longitudinal sound waves. 
Therefore, it is difficult to observe any change at 
$T=T_c$ for the transverse wave, which may explain 
the experimental results. 

Here we neglect the coupling to the collective mode.
Indeed, even in the diffusive limit, $C_{11}$ and $C_{66}$ modes
do couple to the collective mode.
However, our investigation showed that the coupling to the 
collective mode in these cases is almost impossible to detect 
although we do not present the details of the analysis here.

\subsection{Ultrasonic attenuation for all temperature regimes}

The general expression of the sound attenuation coefficient for
$T < T_c$ can be obtained by following the procedure of 
Kadanoff \& Falko, and Tsuneto\cite{kadanoff}. 
%Here we again consider the regime $|{\bf q}| l \ll 1$ where 
%$l$ is the mean free path. 
%We neglect the coupling with the collective mode and take
%into account only the effect of nonmagnetic impurity.
%Let us first consider the attenuation of the transverse sound wave.
To obtain the ultrasonic attenuation coefficient, $\alpha_t$, 
we compute the imaginary part of the correlation function
$\langle [\tau_{xy},\tau_{xy} ] \rangle$.
We finally arrive at
\begin{equation}
{\rm Im} \langle [\tau_{xy},\tau_{xy}] 
\rangle = {p^4_F \over m^2} N(0) \omega \int^{\infty}_{-\infty} 
d\omega \left (-{\partial n_F \over \partial \omega} \right )
{g({\tilde \omega}) \over {\rm Im} \sqrt{{\tilde \omega}^2-\Delta^2}}
\frac{1+y^2/2-\sqrt{1+y^2}}{y^4} \ ,
\end{equation}
where $y=\frac{v_F q}{2{\rm Im} \sqrt{{\tilde \omega}^2-\Delta^2}}$
and $n_F(\omega)=1/(e^{\omega/T}+1)$ is the Fermi distribution 
function. 
The coherence factor $g({\tilde \omega})$ is given by 
\begin{equation}
g({\tilde \omega})=\frac{1}{2}
\left ( 1+\frac{|{\tilde x}|^2-1}{|{\tilde x}^2-1|}
\right ) \ , 
\end{equation}
where
${\tilde x}={\tilde \omega}/\Delta$ is determined from
\begin{equation}
{\tilde x}=\frac{\omega}{\Delta}+i\frac{\Gamma}{\Delta}
\frac{\sqrt{{\tilde x}^2-1}}{{\tilde x}} \ .
\end{equation}
Similar analysis can be also done for $\alpha_l$.

In the limit of $|{\bf q}| l \ll 1$, the above result leads to
the following ratio between the 
attenuation coefficients in the superconducting state, 
$\alpha_{l,t}$, 
and the normal state, $\alpha_{l,t}^n$.
\begin{equation}
{\alpha_{l,t} \over \alpha_{l,t}^n} = \frac{\Gamma}{8\Delta} 
\int_0^{\infty} \frac{d\omega}{T}
{\rm sech}^2 (\frac{\omega}{2 T}) 
\frac{g({\tilde \omega})}{{\rm Im} \sqrt{{\tilde x}^2-1}} \ .
%\frac{1+y^2/2-\sqrt{1+y^2}}{y^4} \ .
\label{att}
\end{equation}
Notice that the Eq. (\ref{att}) applies for both of the 
transverse and longitudinal sound waves. This result is 
evaluated numerically and shown in Fig.2 for
several $\Gamma/\Gamma_c$ where $\Gamma_c = \Delta(0)/2$ 
is the critical scattering rate which drives $T_c$ to zero.

\section{Conclusion}

We have identified a unique collective mode called
the clapping mode in a $p$-wave superconductor with 
the order parameter given by Eq.(1). This collective mode
couples to the sound wave and affects its dynamics. 

The effect of the clapping mode on the sound waves
was calculated in the collisionless limit. 
However, unlike the case of $^3$He, the detection of the 
collective mode appears to be rather difficult.
One needs, at least, the high frequency experiment 
with $\omega \sim  O(1)$ GHz.

In the diffusive limit, we worked out the sound velocity change
near $T=T_c$ and found that it decreases linearly in $1-T/T_c$
which is consistent with the experiment reported by Matsui 
{\it et al} \cite{matsui}.
We also obtained the ultrasonic attenuation coefficient 
for the whole temperature range, which can be tested 
experimentally.
On the other hand, the coupling of the collective mode
is almost invisible in the diffusive limit.

\acknowledgements
We thank T. Ishiguro, K. Nagai, Y. Maeno, and M. Sigrist for helpful 
discussion and especially E. Puchkaryov for drawing Fig.2.
The work of H.-Y. Kee was conducted under the auspices of the Department
of Energy, supported (in part) by funds provided by the University of
California for the conduct of discretionary research by Los Alamos
National Laboratory.
This work was also supported by Alfred P. Sloan Foundation 
Fellowship (Y.B.K.), NSF CAREER Award No. DMR-9983731 (Y.B.K.), 
and CREST (K.M.). 

\appendix

\section{} 
\subsection{Clapping mode and its coupling to the stress tensor}
The fluctuation of the order parameter corresponding to the
clapping mode can be written as
$\delta \Delta \rho_3 \sim e^{\pm 2i\phi} \sigma_1 \rho_3$.
The relevant correlation functions for the couplings are
\begin{eqnarray}
\langle \delta\Delta, {\rm cos}(2\phi) \rangle 
(i\omega_{\nu},{\bf q}) &=& T \sum_n 
\sum_{\bf p} {\rm Tr} [\delta\Delta \rho_3 G({\bf p},\omega_n)
{\rm cos}(2\phi) \rho_3 G({\bf p}-{\bf q},i\omega_n-i\omega_{\nu})] \ , \cr
\langle \delta\Delta, \delta\Delta \rangle 
(i\omega_{\nu},{\bf q}) &=& T \sum_n 
\sum_{\bf p} {\rm Tr} [\delta\Delta\rho_3 G({\bf p},\omega_n)
\delta\Delta\rho_3 G({\bf p}-{\bf q},i\omega_n-i\omega_{\nu})] \ .
\end{eqnarray}
After summing over {\bf p}, we get
\begin{eqnarray}
\langle \delta\Delta, {\rm cos}(2\phi) \rangle &=& 
\left < \pi T N(0) \sum_n \left(\frac{-i\omega_{\nu}\Delta}
{2 \sqrt{\omega_n^2+\Delta^2} \sqrt{\omega_{n+\nu}^2+\Delta^2}} \right)
\frac{\sqrt{\omega_n^2+\Delta^2}+\sqrt{\omega_{n+\nu}^2+\Delta^2}}
{\left(\sqrt{\omega_n^2+\Delta^2}+\sqrt{\omega_{n+\nu}^2+\Delta^2}\right)^2
+\zeta^2} \right > \ , \cr
\langle \delta \Delta,\delta\Delta \rangle &=& 
\left < \pi T N(0) \sum_n \left(1+ \frac{\omega_n \omega_{n+\nu}}
{\sqrt{\omega_n^2+\Delta^2} \sqrt{\omega_{n+\nu}^2+\Delta^2}} \right)
\frac{\sqrt{\omega_n^2+\Delta^2}+\sqrt{\omega_{n+\nu}^2+\Delta^2}}
{\left(\sqrt{\omega_n^2+\Delta^2}+\sqrt{\omega_{n+\nu}^2+\Delta^2}\right)^2
+\zeta^2} \right > \ ,
\end{eqnarray}
where $\langle \cdots \rangle$ on the right hand side of the equations
represents the angle average.
Now summing over $\omega_n$ and analytic continuation 
$i \omega_{\nu} \rightarrow \omega + i \delta$ lead to
\begin{eqnarray}
\langle \delta \Delta, {\rm cos}(2\phi) \rangle
&=& N(0) \langle \frac{\omega}{4\Delta} F \rangle,
\nonumber\\
\langle \delta\Delta, \delta \Delta \rangle
&=& g^{-1} - N(0) \langle \frac{\zeta^2+2\Delta^2-\omega^2}
{4\Delta^2} F \rangle \ ,
\end{eqnarray}
where $F$ is given by
\begin{equation}
F (\omega,\zeta) = 4 \Delta^2 (\zeta^2-\omega^2) \int_{\Delta}^{\infty}
dE \frac{\tanh{(E/2T)}}{\sqrt{E^2-\Delta^2}}
\frac{(\zeta^2-\omega^2)^2-4 E^2(\omega^2+\zeta^2)+4 \zeta^2 \Delta^2}
{[(\zeta^2-\omega^2)^2+4 E^2(\omega^2-\zeta^2)+4 \zeta^2 \Delta^2]^2-16
\omega^2 E^2 (\zeta^2-\omega^2)^2} \ .
\end{equation}

In the limit of $\omega \ll v_F |{\bf q}|$, the contribution 
(in $\langle [h_l,h_l] \rangle$; see Appendix B 3)
due to the coupling with the clapping mode becomes
\begin{eqnarray}
\frac{\langle \delta\Delta, {\rm cos}(2\phi) \rangle^2}
{g^{-1}-\langle \delta\Delta, \delta\Delta \rangle}
&=& N(0) \frac{\omega^2 \langle F \rangle^2}
{4 \langle (\zeta^2 - 2\Delta^2 - \omega^2) F \rangle}\nonumber\\
&\approx& N(0) 
\frac{s^2 f}{4[{1 \over 2}-(2\Delta/v_F q)^2-s^2]} \ . 
\end{eqnarray}
where $f = {\rm lim}_{q \rightarrow 0} 
{\rm lim}_{\omega \rightarrow 0} \langle F \rangle$ is
the superfluid density and given by Eq.(\ref{f}).
We can see that the frequency of the clapping mode is
given by $\sqrt{2}\Delta$ from 
$\langle \delta\Delta, \delta\Delta \rangle$.

\section{}
%\subsection{Stress tensor and its autocorrelation function}
%
%The autocorrelation function,
%$\langle [\tau_{ij}, \tau_{ij}] \rangle$,
%of the stress tensor
%$\tau_{ij}$ is defined as
%\begin{equation}
%\langle [\tau_{ij}, \tau_{ij}] \rangle ({\bf r}-{\bf r'}, t-t')
%\equiv - i \theta (t-t')
%\langle [\tau_{ij} ({\bf r},t), \tau_{ij} ({\bf r'},t') ]
%\rangle \ ,
%\end{equation}
%where
%\begin{equation}
%\tau_{ij} ({\bf r},t) = \sum_{\sigma}
%\left [
%{(\nabla-\nabla^{'})_i \over 2i}{(\nabla-\nabla^{'})_j \over
%2im} \psi^{\dagger}_{\sigma} ({\bf r},t)
%\psi_{\sigma} ({\bf r'},t)
%\right ]_{{\bf r'}={\bf r}} \ .
%\end{equation}
%Here $\psi^{\dagger}_{\sigma}$ is the electron creation
%operator with spin $\sigma$.
%The relation between the stress tensor and the dynamics of 
%the sound waves is extensively discussed in the work of
%Kadanoff and Falko.\cite{kadanoff}

\subsection{Longitudinal sound wave in the collisionless limit}
The longitual sound velocity
shift is given by the real part of
$\langle [h_l,h_l] \rangle$.
The irreducible correlation function for the stress tensor
can be obtained from
\begin{eqnarray}
\langle [\tau_{xx},\tau_{xx}] \rangle_{00} =
T \sum_n \sum_{\bf p} {\rm Tr} [{p_F^4 \over m^2}
({\rm cos} \ \phi)^4 \rho_3 G({\bf p},\omega_n)
\rho_3 G({\bf p}-{\bf q},i\omega_n-i\omega_{\nu})] \ ,
\end{eqnarray}
where $\phi$ is the angle between  ${\bf p}$ and ${\bf q}$.
Since the stress tensor couples to the density, the correlation
function is renormalized as
\begin{equation}
\langle [h_l,h_l] \rangle_{0} =
\langle [h_l,h_l] \rangle_{00} +
\frac{V({\bf q}) \langle [h_l,n] \rangle
\langle [n,h_l] \rangle} {1-V({\bf q}) \langle  [n,n] \rangle} ,
\end{equation}
where
$V({\bf q})=2 \pi e^2/|{\bf q}|$ is the
Coulomb interation.
This equation can be simplified in the long wave length limit
($|{\bf q}| \rightarrow 0$) as
\begin{equation}
\langle [h_l,h_l] \rangle_{0} \approx
 \left( \frac{q}{\omega} \right)^2
[ \langle [\tau_{xx},\tau_{xx}] \rangle_{00} -
\frac{ \langle [\tau_{xx},n] \rangle
\langle [n,\tau_{xx}] \rangle} {\langle  [n,n] \rangle}] \ .
\label{sv1}
\end{equation}

It is useful to define the following quantity for notational
convenience.
\begin{equation}
\langle A, B \rangle =
T \sum_n \sum_{\bf p} {\rm Tr} [ A
\rho_3 G({\bf p},\omega_n)
B \rho_3 G({\bf p}-{\bf q},i\omega_n-i\omega_{\nu})] \ ,
\end{equation}
where $A$ and $B$ can be some functions of $\phi$ or
operators.
Using this notation,
Eq.(\ref{sv1}) can be rewritten as
\begin{eqnarray}
\langle [h_l,h_l] \rangle_{0} &\approx&
\left( \frac{q}{\omega} \right)^2 {p^4_F \over m^2}
\left [ \langle {\rm cos}^2 \phi, {\rm cos}^2 \phi \rangle -
{\langle {\rm cos}^2 \phi, 1 \rangle
\langle 1, {\rm cos}^2 \phi \rangle \over
\langle 1, 1 \rangle} \right ] \cr
&=&\left( \frac{q}{\omega} \right)^2  {p^4_F \over 4 m^2}
\left [ \langle {\rm cos}(2\phi), {\rm cos}(2\phi) \rangle
- {\langle {\rm cos}(2\phi), 1 \rangle
\langle 1, {\rm cos}(2\phi) \rangle \over
\langle 1, 1 \rangle} \right ] \ .
\label{sv2}
\end{eqnarray}

Then,
each correlation function can be computed from
\begin{eqnarray}
\langle 1, 1 \rangle (i\omega_{\nu},{\bf q}) &=& T \sum_n
\sum_{\bf p} {\rm Tr} [\rho_3 G({\bf p},\omega_n)
\rho_3 G({\bf p}-{\bf q},i\omega_n-i\omega_{\nu})] \ , \cr
\langle 1, {\rm cos}(2\phi) \rangle (i\omega_{\nu},{\bf q}) &=& T \sum_n
\sum_{\bf p} {\rm Tr} [{\rm cos}(2\phi) \rho_3 G({\bf p},\omega_n)
\rho_3 G({\bf p}-{\bf q},i\omega_n-i\omega_{\nu})] \ , \cr
\langle {\rm cos}(2\phi), {\rm cos}(2\phi)
\rangle (i\omega_{\nu},{\bf q}) &=& T \sum_n
\sum_{\bf p} {\rm Tr} [{\rm cos}^2(2\phi) \rho_3 \sigma_1
G({\bf p},\omega_n)
\rho_3 G({\bf p}-{\bf q},i\omega_n-i\omega_{\nu})] \ .
\label{tautau}
\end{eqnarray}

Summing over {\bf p} leads to
\begin{equation}
\langle 1, 1 \rangle
= \left \langle
\pi T N(0) \sum_n \left(1- \frac{\omega_n \omega_{n+\nu}+\Delta^2}
{\sqrt{\omega_n^2+\Delta^2} \sqrt{\omega_{n+\nu}^2+\Delta^2}} \right)
\frac{\sqrt{\omega_n^2+\Delta^2}+\sqrt{\omega_{n+\nu}^2+\Delta^2}}
{\left(\sqrt{\omega_n^2+\Delta^2}+\sqrt{\omega_{n+\nu}^2+\Delta^2}\right)^2
+\zeta^2} \right \rangle \ ,
\label{complex}
\end{equation}
where $\zeta = {\bf v}_F \cdot {\bf q}$ and $N(0)=m/2\pi$ is
the two dimensional density of states.
Similar equations can obtained for
$\langle {\rm cos}(2\phi), 1 \rangle$ and
$\langle {\rm cos}(2\phi), {\rm cos}(2\phi) \rangle$ with
additional angle factors, ${\rm cos}(2\phi)$ and
${\rm cos}^2(2\phi)$ respectively.
After summing over $\omega_n$ and analytic continuation
$i \omega_{\nu} \rightarrow \omega + i \delta$, we get
the following results in the limit of $\omega \ll v_F |{\bf q}|$.
\begin{eqnarray}
\langle 1, 1 \rangle &=& N(0) \langle \frac{\zeta^2-\omega^2 f}
{\zeta^2-(\omega+i\delta)^2} \rangle
\approx N(0) \left [
1-i\frac{s(1-f)}{\sqrt{1-s^2}} \right ] \ , \nonumber\\
\langle {\rm cos}(2\phi), 1 \rangle &=&
N(0) \langle {\rm cos}(2\phi) \frac{\zeta^2-\omega^2 f}
{\zeta^2-(\omega+i\delta)^2} \rangle
\approx N(0) \left [
2 s^2 (1-f)(1+i \frac{1-2s^2}{2s\sqrt{1-s^2}})
\right ] \ , \nonumber\\
\langle {\rm cos}(2\phi), {\rm cos}(2\phi) \rangle &=&
N(0) \langle {\rm cos}^2(2\phi)
\frac{\zeta^2-\omega^2 f} {\zeta^2-(\omega+i\delta)^2} \rangle
\approx N(0) \left [
\frac{1}{2}-2s^2(1-f)-i\frac{(1-f)s}{2\sqrt{1-s^2}} \right ] \ .
\end{eqnarray}

We find that the second term in the last line of Eq.(\ref{sv2})
is of higher order in $ \omega / v_F |{\bf q}| (\equiv s)$ so that
we can ignore it. In Sr$_2$RuO$_4$ or metals, $s \ll 1$.
Thus the effect of the coupling to the density is merely to
change the vertex associated with $\tau_{xx}$ from
${p_F^2 \over m} {\rm cos}^2\phi$ to
${p_F^2 \over 2m} {\rm cos}(2\phi)$ as far as the lowest
order contribution is concerned.
Evaluation of $\langle {\rm cos}(2\phi), {\rm cos}(2\phi) \rangle$
leads to
\begin{equation}
{\rm Re} \langle [h_l,h_l] \rangle_{0} \approx
\left( \frac{q}{\omega} \right)^2 {\rm Re} {p^4_F \over 4 m^2}
\langle {\rm cos}(2\phi), {\rm cos}(2\phi) \rangle
= \left( \frac{q}{\omega} \right)^2 {p^4_F \over 4 m^2} N(0) \left [
\frac{1}{2}-2 s^2 (1-f)
\right ] \ ,
\end{equation}
where $f$ is the superfluid density.

\subsection{Contribution coming from the coupling 
to the clapping mode}
The correction due to the collective mode leads to the renormalized
correlation function as follows.
\begin{equation}
\langle [h_l,h_l] \rangle =
\langle [h_l,h_l] \rangle_{0}
+ \frac{\langle [h_l,\delta \Delta \rho_3] \rangle
\langle [\delta\Delta \rho_3,h_l] \rangle}
{g^{-1} - \langle [\delta \Delta \rho_3,\delta \Delta \rho_3]
\rangle} \ ,
\end{equation}
where $g$ is the coupling constant between the stress tensor
and the collective mode. $\delta \Delta \rho_3$ represents the
fluctuation associated with the clapping mode.
Using the fact that $\langle 1, e^{2i\phi} \rangle = 0$ and
$\delta \Delta \sim e^{2i\phi} \sigma_1$, the above
equation can be further reduced to
\begin{eqnarray}
\langle [h_l,h_l] \rangle &=&
\left( \frac{q}{\omega} \right)^2 {p^4_F \over 4 m^2} \left [
\langle {\rm cos}(2\phi), {\rm cos}(2\phi) \rangle
+ \frac{\langle {\rm cos}(2\phi),\delta \Delta \rangle
\langle \delta \Delta, {\rm cos}(2\phi) \rangle}
{g^{-1} - \langle \delta \Delta, \delta \Delta
\rangle} \right ] \ \nonumber\\
&=&
\left( \frac{q}{\omega} \right)^2 {p^4_F \over 4 m^2} \left [
\frac{1}{2}-2
\left ({C_l \over v_F} \right )^2 \left ( 1-f-
\frac{f (v_F |{\bf q}|)^2}
{4 \{(v_F|{\bf q}|)^2+4\Delta(T)^2-2\omega^2\} } \right ) \right ]
+ i {m_{\rm ion} C_l \over \omega} \alpha_l (\omega) \ ,
\label{recorr}
\end{eqnarray}
where $f$ is the superfluid density
and given by Eq.(\ref{f}).

\begin{figure}    
%\hspace{1.0truecm}    
\vspace{-2.5truecm}
\center    
\centerline{\epsfysize=8.0in    
\epsfbox{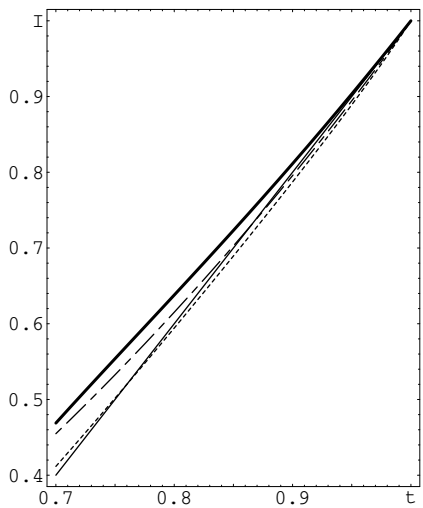}}
\vspace{-12.0truecm}    
\begin{minipage}[t]{10.0cm}    
\caption{The function $I$ representing the reduction in
the sound velocity as a function of the reduced 
temperature $t=T/T_c$ with $0.7 < t < 1.0$ for 
$v_F|{\bf q}| / \Delta(0) = 0, 1, 2, 3$.}     
\end{minipage}    
\end{figure}    

\begin{figure}    
%\hspace{1.0truecm}    
\vspace{-2.5truecm}
\center    
\centerline{\epsfysize=8.0in    
\epsfbox{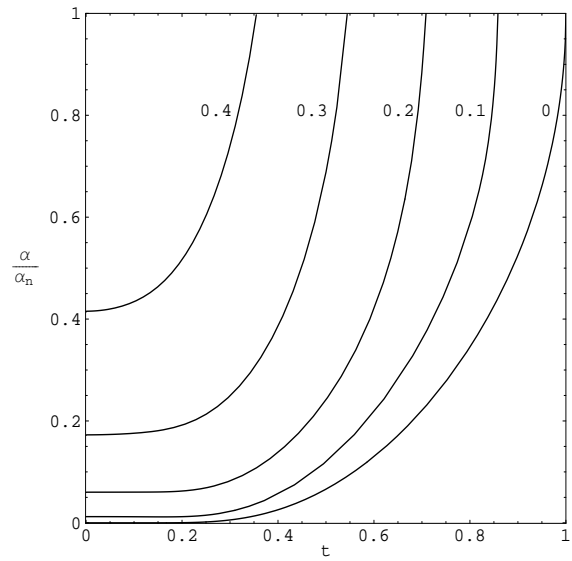}}
\vspace{-10.5truecm}    
\begin{minipage}[t]{10.0cm}    
\caption{The normalized attenuation coefficient as a function
of the reduced temperature $t=T/T_c$ for 
$\Gamma/\Gamma_c = 0, 0.1, 0.2, 0.3, 0.4$.}     
\end{minipage}    
\end{figure}    

\end{document}